# *Part* I. *Relativistic "Paradoxes"*

## Ch. 1. Physics of the Relativistic Length Contraction and the Ehrenfest Paradox

Moses Fayngold
*Department of Physics, New Jersey Institute of Technology, Newark, NJ 07102*

Relativistic kinematics is usually considered only as a manifestation of pseudo-Euclidean (Lorentzian) geometry of space-time. However, as it is explicitly stated in General Relativity, the geometry itself depends on dynamics - specifically, on the energy-momentum tensor.

We discuss a few examples, which illustrate the dynamical aspect of the length-contraction effect within the framework of Special Relativity.

We show some pitfalls associated with direct application of the length contraction formula in cases when an extended object is accelerated. Our analysis reveals intimate connections between length contraction and the dynamics of internal forces within the accelerated system.

The developed approach is used to analyze the correlation between two *congruent* disks - one stationary and one rotating (the Ehrenfest paradox). Specifically, we consider the *transition* of a disk from the state of rest to a spinning state under the applied forces. It reveals the underlying physical mechanism in the accompanying transform of Euclidean geometry of stationary disk to Lobachevsky's (hyperbolic) geometry of the spinning disk in the process of its rotational boost. A conclusion is made that the rest mass of a spinning disk or ring of a fixed radius must contain an additional term representing the potential energy of non-Euclidean circumferential deformation of its material. Possible experimentally observable manifestations of Lobachevsky's geometry of rotating systems are discussed.

## Introduction

The dominating view inferred from the so-called *relativistic kinematic effects* (length contraction and time dilation) is that these effects are exclusively manifestations of Lorentzian geometry of Minkowski's space-time [1, 2]. This axiomatic approach is elegant and efficient in cases of uniform motion, but fails when the state of motion is changing. Study of these much more general cases provides a far deeper insight into the nature of "kinematic effects". It shows that emphasizing Minkowski's geometry as the single basis for understanding relativistic kinematics oversimplifies the actual physics. It would also go against the spirit of the General Relativity, according to which space-time geometry itself is determined by the energy-momentum tensor and the corresponding boundary conditions [2-8].

A close look at relativistic kinematics shows that it also has dynamical underpinnings, and moreover, the "embryo" of the intimate connection between geometry and matter can be traced already in Special Relativity (SR).

The dynamical aspects of relativistic kinematics were explicitly formulated at least as far back as 1975 [9]. The dynamics in kinematic effects can be clearly seen in accelerated motions. One can come across the statements that SR is not applicable to accelerated motions (see, e.g. [10]). This is a widely spread misconception which contributes to the over-simplistic view of kinematic effects in SR.

Relativistic motion of accelerated objects has been extensively discussed (see, e.g., [11-16]), including conditions with time-independent distributed external forces and a time-dependent local force. In [17, 18] the motion of a train under the time-dependent distributed force and the corresponding physical mechanisms associated with Lorentz-contraction effect have been analyzed.

In this article we also generalize the approximation of instant forces to more realistic cases of finite forces acting during finite time intervals.

Apart from comparing the length measurements carried out by different inertial observers [11-13, 17, 18], we consider two different accelerating mechanisms (friction and jet propulsion), and study what actually happens to an object when its motion is rapidly changed. We show that the accelerating program preserving the proper length of an object as described in [19] is not the only one possible. We discuss the role of atomic interactions within the object in determining its final state after removal of the external forces.

In the second part we apply the same approach to study the dynamics of "kinematic" processes during a rotational boost of a disk. As is well known, the spinning disk is in a state of a complex deformation and its spatial geometry is non-Euclidean [7], which explains the Ehrenfest paradox. Our approach unveils the underlying physics in the *transition* from the Euclidian geometry of a stationary disk to the hyperbolic (Lobachevsky's) geometry of rotating disk and shows its connection with the fact that global time in a rotating system is not single-valued. The change of geometry on the spinning disk is accompanied by the corresponding dynamical effect: in any rotational boost an additional energy must be spent on circumferential deformation of every annulus of the disk. Accordingly, the rest mass of the spinning disk acquires an additional term representing potential energy of this deformation. In the non-rotating rest frame this addition may be small relative to the increase of relativistic mass of the disk. But in the co-rotating frame it will be (assuming no radial changes) the only remaining additional term. This term may be essential in rapidly rotating relativistic objects like neutron stars.

## 1. The dynamics of relativistic length contraction

Let a horizontal rod of proper length $L_0$ pass with a speed $V$ through a barn of length $l_0 < L_0$ [1]. Suppose the ratio $L_0 / l_0$ just equals the Lorentz-factor $\gamma(V) \equiv \left(1-(V/c)^2\right)^{-1/2}$ determining the Lorentz- contracted length of the rod. Consider two *inertial* reference frames: one associated with the barn (system A), and the other *originally* associated with the rod (system B). Accordingly, we introduce two observers: Alice in A, and Bob in B. Due to our assumption, the *moving* rod's length measured in A is

$$\frac{L_0}{\gamma(V)} = l_0 \tag{1}$$

Then at a certain moment the rod will exactly fit into the barn. Suppose right at this moment we stop the rod. A formal application of the rule (1) might lead one to expect that once the rod is stopped ($V = 0$) its length becomes $L_0$ due to disappearance of length contraction. If so, its edges would now stick by a distance $(L_0 - l_0)/2$ out of the barn (Fig. 1).

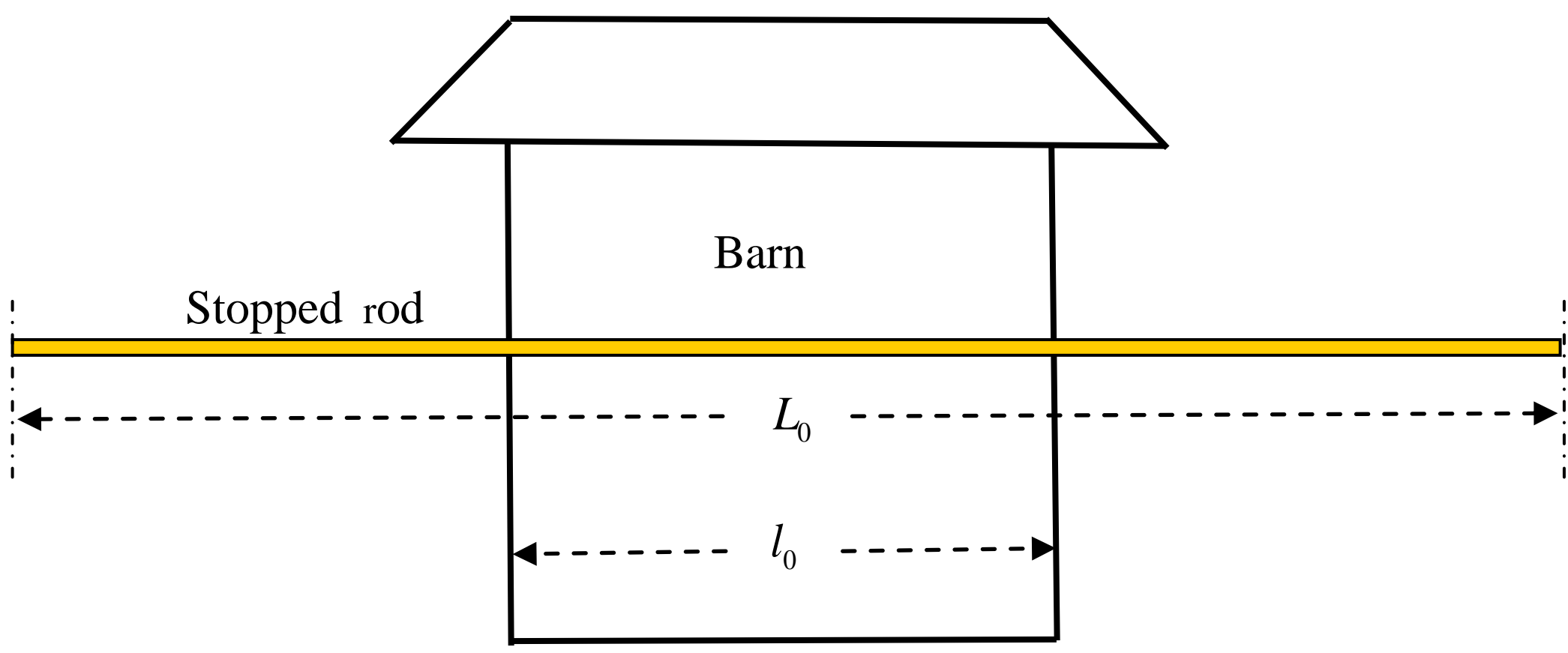

**Fig. 1**. Stopped rod as naively expected only from the fact that it is now at rest in A.

In reality, however, such outcome is impossible in the described situation. The actual result will critically depend on stopping procedure and structure of the rod. We will analyze four different acceleration programs (I) – (IV).

(I) *Stopping the rod simultaneously in* A *as described in the beginning*.

Suppose the braking forces are dominating and are applied to all parts of the rod *simultaneously in A*. As a possible model, consider a very sticky horizontal shelf approaching from beneath (Fig. 2) and touching the rod right at the moment when it is within the barn. Immediately, equal friction forces are applied to equal parts of the rod (assume that produced heat quickly dissipates). In this case, the outcome shown in Fig.1 would contradict the dynamics of motion. Under conditions, all parts of the rod will slow down synchronously in A, so the length of the rod cannot change. When stopped, it remains the same as it was in motion, which means that its *proper* length has decreased from $L_0$ to $l_0$. This means, the stopped rod is physically compressed. The specific mechanism responsible for it is associated with relativity of simultaneity. According to it, the local forces applied to different parts of the rod simultaneously in A, are not simultaneous in B [17, 18]. In Fig. 2, at the zero moment $t_l = t_r = 0$ of Alice's time (when the shelf touches the rod and stops it), Bob's synchronized clocks at the rod's edges read, in the chosen units, $t_l' = -1$ at its right and $t_r' = 1$ at its left edge. Accordingly, the rod stops first at its front. This compresses the rod! The assumption of the rod slowing down simultaneously in all its parts in A is incompatible with preserving the rod's *proper* length during the stop. The *proper* length does not generally conserve with change of object's motion.

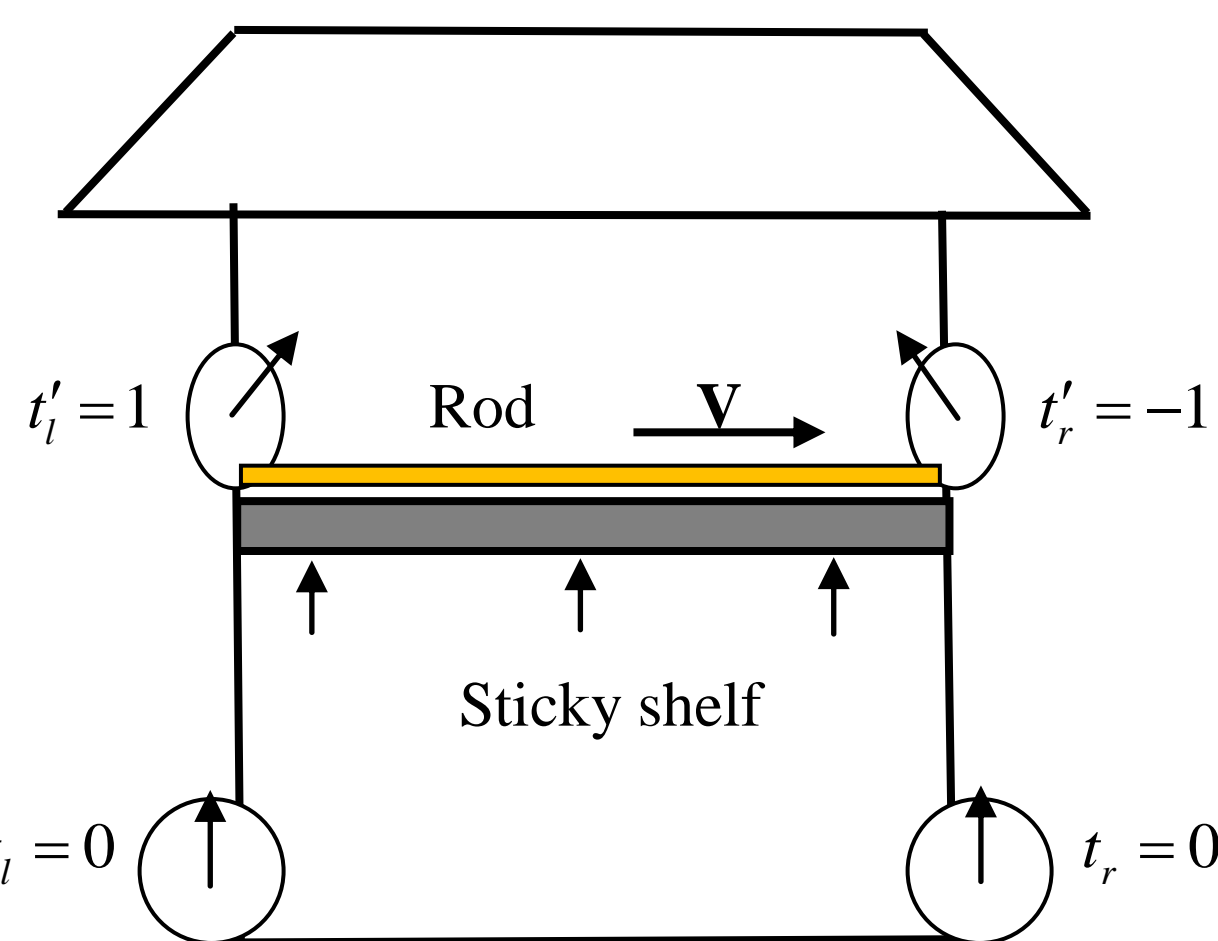

**Fig. 2.** The rod stopped by a friction with the sticky shelf approaching it from beneath. Shown is the moment right before the stop, when the rod's edges coincide with the opposite sides of the barn. At this moment, the clearing between rod and the shelf disappears and all Alice's clocks record $t = 0$. Bob's clocks on the rod read different times for the same events.

The case at hand provides a specific illustration. Here the *disappearance* of length contraction due to the stop is counterbalanced by the opposing *dynamic compression* (actual physical deformation). Accordingly, while Alice sees no change in the length of the rod, Bob observes its dramatic decrease. SR predicts this change and explains it as the result of relativity of simultaneity. For detailed explanation, we turn to Bob's account of the process.

Bob sees the barn sliding to the left along the rod. The barn's length in B is Lorentz-contracted to $l = l_0 / \gamma = L_0 / \gamma^2$, which is *less* than length of the rod. While the rod's edges touch the opposite sides of the barn simultaneously in A, these events happen at different moments in B. The barn's back door reaches the right edge of the rod at $t_r' = -1$ of Bob's time. The front door coincides with the left edge at $t_l' = 1$. The time interval between the events is 2 units. Applying the Lorentz transformation for time coordinates of the two events, taking into account that these events happen both at $t = 0$ in A, and using (1), we can express this time interval in terms of the initial proper length of the rod:

$$\Delta t' = \gamma(V)\,\frac{Vl_0}{c^2} = \frac{VL_0}{c^2}\ . \tag{2}$$

The non-zero time discrepancy (2) seems to clash with horizontality of sticky shelf stopping the rod, but it does not. Rising shelf horizontal in A is tilted in B (Fig.3).

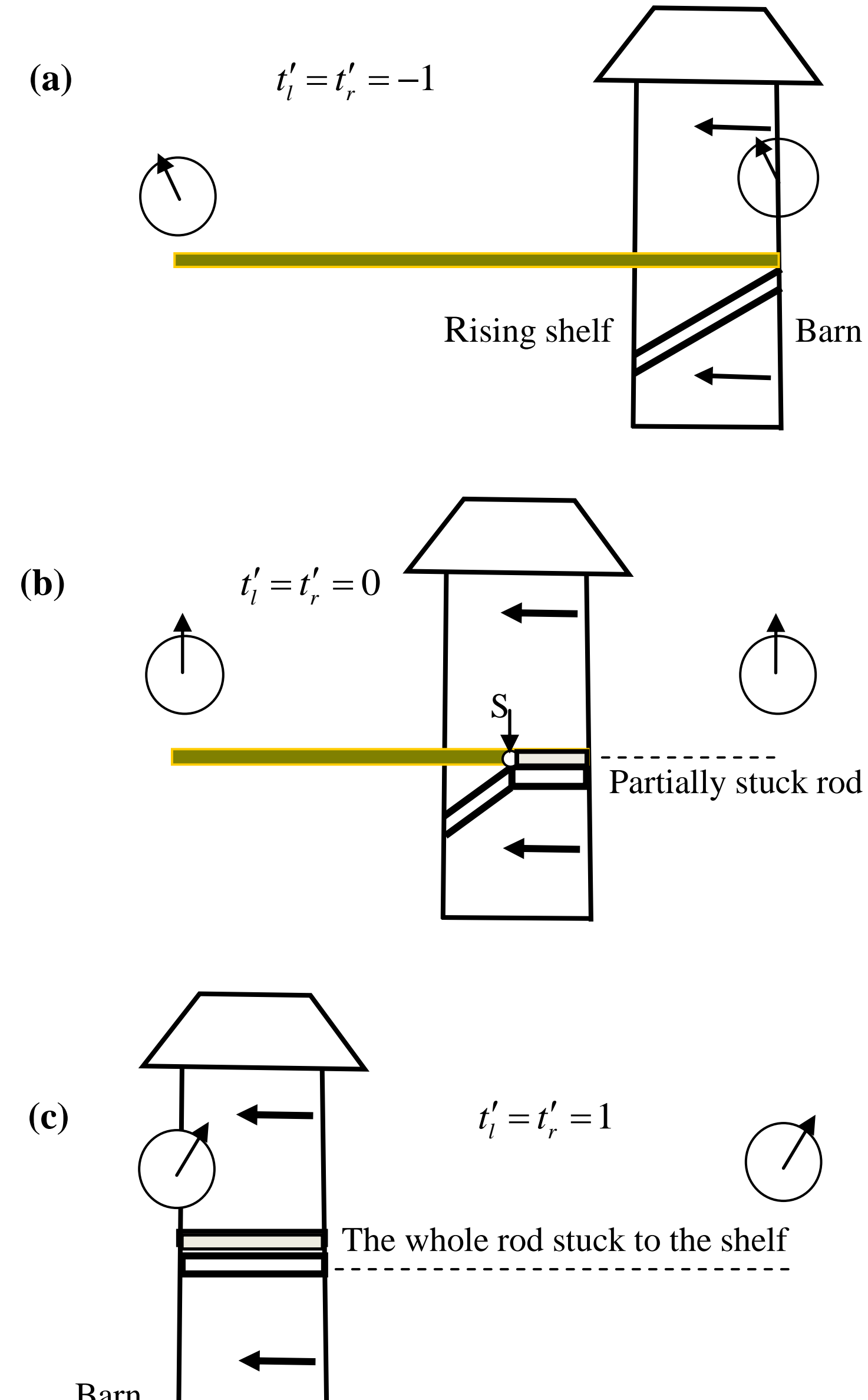

**Fig. 3**. Different stages of stopping the rod as observed by Bob.
The rising (and accordingly tilted) sticky shelf involves in motion the adjacent parts of the rod. S is the running point between horizontal and tilted parts of the shelf. It separates the rod into compressed and relaxed parts. In the end of the process the whole rod is compressed and moving together with the barn.

If the edges of the shelf are recorded *simultaneously in B*, these records have different moments in A – the one at the rear edge is later (when the shelf is accordingly higher) than one at the front. Therefore the whole shelf observed at one moment in B, has its trailing edge higher than its front edge by a certain amount $\Delta y'$. Any rising surface horizontal in A has a tilt angle $\alpha$ in B given by [1, 17, 18]:

$$\tan\alpha = \frac{\Delta y'}{l} = \gamma \frac{V v_\perp}{c^2} = \gamma^2 \frac{V v'_\perp}{c^2} \qquad (3)$$

Here $v_\perp$ is the shelf's velocity in A and $v'_\perp = v_\perp / \gamma(V)$ is the vertical component of its velocity in B. At $t'_r = -1$ of Bob's time, the upper end of the shelf touches the right edge of the rod, sticks to it and drags it to the left (Fig. 3a). As mentioned above, this starts compressing the rod. As the rest of the shelf continues to rise, there forms region of compression rapidly expanding to the left along the rod due to joining of its originally relaxed parts to the shelf (Fig. 3b). The front of the expanding region is the running separation point (denote it *S*) between the two parts: the one to its left – still intact, and the one to the right already moving together with the barn. Point *S* sweeps down the rod faster than light. This kind of superluminal motion does not contradict anything [17, 18, 20-25]; it is just a moving boundary between two regions rather than a physical particle. Particles of the compressed segment of the rod are moving (together with the corresponding part of the shelf) slower than light. Point *S* outruns them because, as mentioned above, the new particles in front join this motion as they grip with the rising part of the shelf. The process ends up with the whole rod stuck to the shelf and contracted to the size of the barn (Fig. 3c).

The expanding of compressed part here is *not* the shock wave. In the latter, the motion of particles right behind the wave front is the cause, and the motion of the particles right ahead will be the effect. In the expanding compression described here, any particle of the rod changes its motion only due to the contact with the rising part of the shelf – before it knows anything about any changes in the adjacent part of the rod. The changes of state on two sides of *S* are not in the "cause and effect" relation. If the compression were to produce a shock wave, the latter would also lag behind *S*, since any shock wave is subliminal. Such a wave does not form here – the shelf freezes the corresponding motion.

During this process we can no longer consider the rod as *one* inertial reference frame (RF). It belongs to two different frames: one to the left of the advancing front is still part of system B, and the one to the right of the front is already part of A. The separation into two systems starts from the whole rod forming one system B, and ends up with the whole rod belonging (together with the barn) to frame A. During the mixed "affiliation" between these two events, the rod *cannot be characterized by a certain proper length* in its conventional definition as the length of a stationary object in its rest frame. In the described process, the two parts of the rod have different rest frames. In such case we generalize definition of proper length as the object's length in a RF where its *net momentum* is zero. This RF does not coincide with either A or B but transfers from one to another during the process. Motion of an accelerated object can be described in consecutive steps using the co-moving inertial frames, each with a small speed relative to the previous one, and in each such frame the rod with the zero *net* momentum consists of the two *moving* parts whose respective lengths change from frame to frame [18, 19]. Actually, this involves a continuous set of inertial frames between A and B. And the respective momentum of each part of the rod changes with time as the $S$- point slides down the rod and we accordingly switch from one RF to another. As a result, the generalized single RF (call it system C) turns out to be accelerating. This should come as no surprise once the whole rod is accelerating from B to A. Therefore we can no longer expect the *generalized* proper length to remain constant. And its definition allows one to watch its change as a function of time.

Using Eq. (2) gives the speed of point $S$ in B:

$$u = \frac{L_0}{\Delta t'} = \frac{c^2}{V} \qquad (4)$$

One can also derive this equation by considering motion of the tilted part of the shelf. Its velocity has two components: $v'_\perp$ (vertical) and $V$ (horizontal). With the first component alone, point $S$ would slide along the rod with the speed $v'_\perp / \tan\alpha = \gamma^{-2} c^2 / V$ . With the second component alone, it would slide along the rod with the speed $V$. The total speed of $S$ is the sum of the two contributions and yields expression (4). Such sum of two collinear motions of a point in one RF should not be confused with Lorentz-transformation for velocity of an object in A when observed in B.

The described "shelf-induced" contraction stops when the rod shrinks down to the size of the barn. The contraction is caused by deformation of the material of the rod under external force.

The presented arguments, evident to Bob, may appear false to Alice. Indeed, all parts of the rod are stopped simultaneously in her barn. The distances between any two adjacent atoms did not change. Therefore it seems there cannot be any change in the inter-atomic interactions, so with heat removed, the rod must remain in its relaxed state.

But if Alice takes a closer look at her experiment she will realize that even with heat instantly removed, what had been a regular rod is now a streak of some exotic super-dense material glued to the shelf. We can understand this by turning to hitherto neglected atomic interactions and their role in the whole process including kinematic effect described by (1).

It is true that inter-atomic distances conserve in A during the braking. But this does *not* mean conservation of the inter-atomic forces. In relativistic dynamics, the interactions between two particles depend also on their velocity [7, 26]. If the latter changes, the interaction force can also change even with the inter-atomic distance fixed.

The *proper* distance between two neighboring atoms in a rod is about $a_0 = 10^{-10} m$. At this distance, each atom is in stable equilibrium (or vibrates around its equilibrium position). We can model such state by a linear chain of equidistant particles connected by springs (Fig. 4a). In equilibrium, the springs are relaxed.

An equilibrium state of a system must be the same in any inertial RF. Alice and Bob must each observe such a state as a stationary chain of atoms separated by the same distance $a_0$. Bob initially has a rod in such state. But Alice observes from her barn the respective inter-atomic distance to be only $a_0 / \gamma(V)$ (Fig. 4b). This is still consistent with equilibrium since the rod is moving. However, if the rod is stopped *without changing its length* in A, she gets a *stationary* object (Fig. 4c) for which the distance $a_0 / \gamma(V)$ is not consistent with equilibrium state. Its atoms are closely packed together. The springs representing interaction forces are compressed to $\gamma^{-1}(V)$ of their normal length. This creates inter-atomic repulsion. In a model representing atomic forces by springs obeying the Hooke's law, sudden removal of the shelf (disappearance of external forces) might send the system into vibrations around its natural *proper* size $L_0 = Na_0$ where *N* is the number of atoms along the chain. If the vibrations are damped, the system will ultimately relax, thus restoring its proper length.

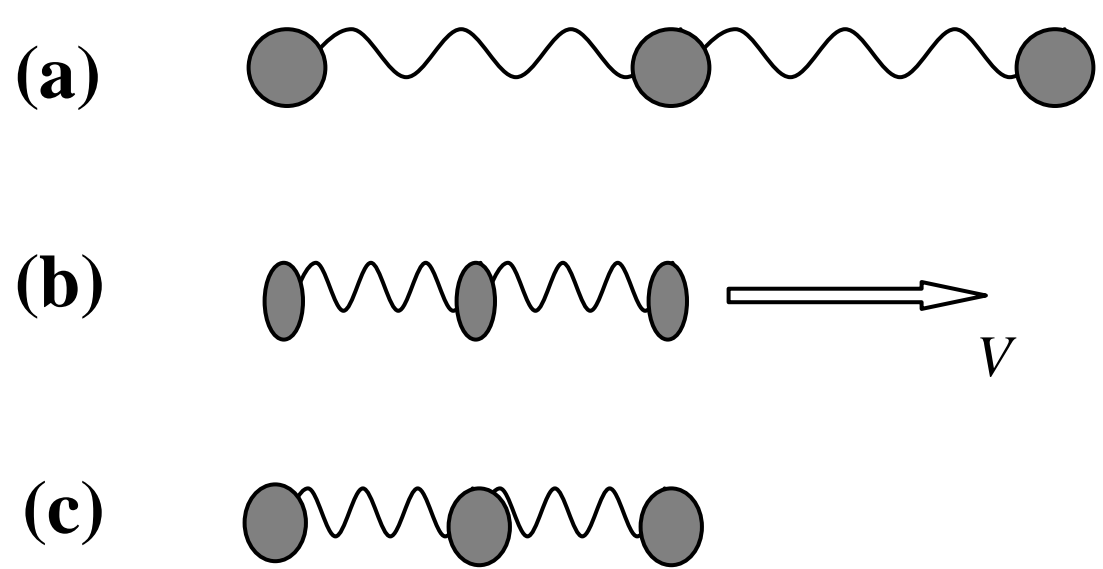

**Fig. 4**. Chain of spheres connected by springs
(a) A chain in stable equilibrium in its rest frame B. The relaxed springs determine distance between neighboring spheres as an intrinsic physical characteristic of the chain.
(b) *The same relaxed* state as observed from A. The chain and the spheres are length-contracted.
(c) The same chain, *stopped* in A by simultaneous forces. The chain is no longer in equilibrium – the *proper* distance between the spheres is reduced, the springs are compressed.

The springs only approximately model internal forces, but the fact remains that these forces act as the "keeper" of proper length. The field of a moving point charge is flattened in the direction of motion [7, 26]. This flattening is an intrinsic property of the field as such and can be considered as its "length contraction". But when the charge stops, its field restores its spherical symmetry (Fig.5).

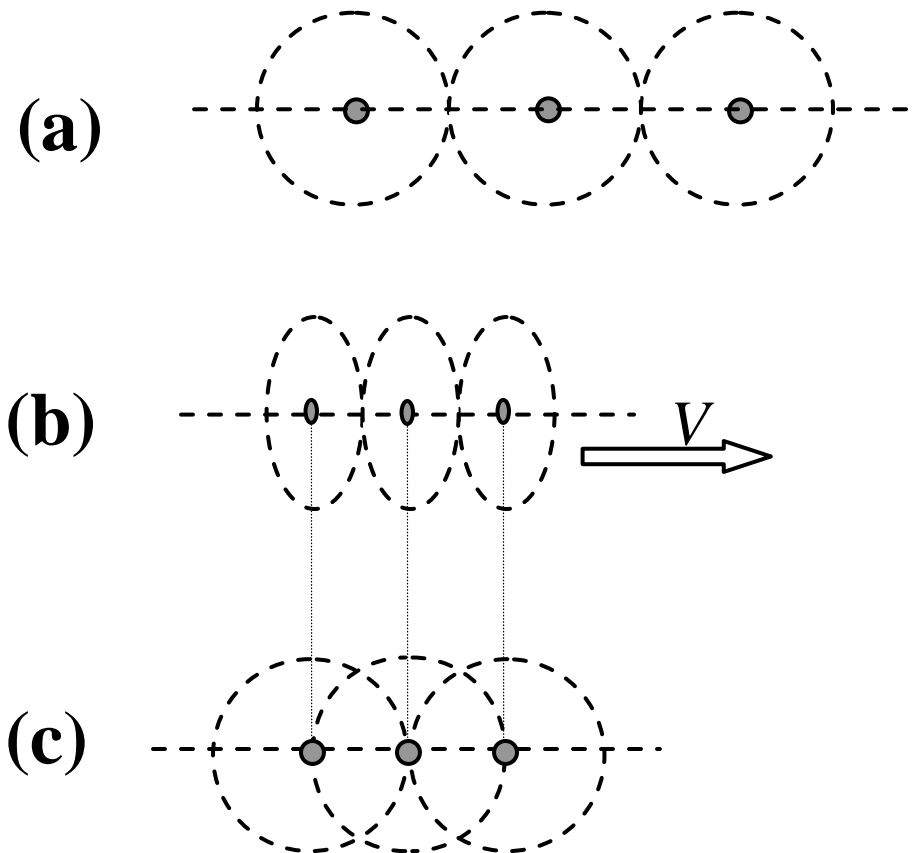

**Fig. 5**. "Equipotential surfaces" of individual electric fields produced by equidistant point charges.
(a) In the initial RF of the system (frame B). The charges are kept in equilibrium by inner forces
(b) The same state as seen in frame A before the system is stopped
(c) In frame A after the stop. The system can be maintained in such state only by external forces.

The same is true for all other forces (and quantum-mechanical probability distributions for all particles), since they all obey the relativistic equations which are Lorentz-invariant. For a moving system, its particles and the *distances* between them must be all Lorentz-contracted along the direction of motion [17, 18, 26-28]. When the rod stops, the field of each of its particles restores its stationary shape and the corresponding inter-atomic distances must accordingly increase. If the particles are not allowed to shift apart, their fields get overlapped (Fig. 5c), which produces strong repulsion. It may blow up the rod if the external forces are removed. Otherwise, we will have the "post-stop" picture observed by Alice, - the rod not kinematically contracted but dynamically compressed. However, if external forces slowly weaken, the inner repulsion can restore the *initial* proper length of the rod. In this respect, *the internal fields act as a memory*, keeping information about object's proper shape.

We can summarize this part in the following way. Before the stop, the rod, albeit length-contracted, was *not* deformed because the inter-atomic distances matched the shape of the internal fields of *moving* atoms. After the stop, even though the length of the rod did not change in A, the rod *is* deformed because the atomic distances no longer match the shape of the *stationary* inner field. The rod was not allowed to readjust to this changed shape. Our concept of deformation should be refined to adequately describe this relativistic effect. Generally, there is no one-to-one correspondence between deforming forces and *change in length* in a given RF.

According to SR, nothing can be ideally rigid [7, 28]. A rod rigid in all conventional situations, behaves as a stick of putty in the above thought experiment. In cases with huge but short-lasting external forces we can use the model of a super-deformable rod and represent it as its end masses connected by a spring. Then we can find how the removal of external forces changes the resulting state.

The work of the external forces is associated with the corresponding energy. In considered case the initial kinetic energy $K$ of moving rod cannot be converted entirely into heat $Q$ at the moment of stop. Right after the stop part of this energy will be found in the form of potential energy $U$ of deformation. If the rod's structure (e.g., linear chain if Fig. 4) renders it sufficiently elastic, we can approximate the corresponding deformation by Hooke's law. The amount $U$ in the thermodynamic relation $K = Q + U$ can then be evaluated as

$$U \cong \frac{1}{2} k L_0^2 \left(1-\gamma^{-1}(V)\right)^2 , \qquad (5)$$

where $k$ is the effective spring constant in the used model. This amount will naturally be incorporated into the rest mass of the stopped rod. Thus, the decrease in proper length will be accompanied by the corresponding increase of the rest mass

$$M_0^{In} \to M_0^{Fin} = M_0^{In} + U/c^2 \qquad (6)$$

If the external forces keeping the compressed state are removed sufficiently slowly to avoid possible explosion, the rod can restore it initial rest mass and proper length.

Altogether, the whole phenomenon, while looking differently in different RF, is consistent for all and predicts a final state on which everybody agrees. To Alice, the rod (right after the stop) retains its initial size because the *equal* external forces are applied *simultaneously* to its equal parts. Such forces stop the rod without changing its length. To Bob, the rod has been compressed because the external forces act *at different times* on different parts of the rod. Such forces both deform the rod *and* transfer it to the passing barn. Both agree that in the final state the rod is physically deformed and just fits into the barn.

(II) *Boosting simultaneously in* B (*accelerating stationary rod*)

Stopping the rod in A means, as before, its boost to the left in B, hence the name of this case. But the boost here will be different from case (I). Now we will act on the rod by local forces *simultaneously in B*. On the face of it, boosting a rod *from its rest frame* may seem a natural way to conserve its proper length. But upon some thought we immediately see this will also fail. Conserved length after boost means, in view of (1.1), extended *proper* length!

It may be instructive to study the process in details. We start with its picture as observed by Bob. As mentioned above, now we can represent the rod by two equal masses at its edges, boosted by the identical jet engines (see also [12, 13]). Assume Bob's initial position in the middle between the masses. When the barn passes by both engines fire to accelerate the rod in the same direction (Fig. 6). Since engines' actions are simultaneous in B, they boost the rod to the left without changing its length. If the engines are sufficiently powerful, the rod instantly acquires the speed of the barn. In the final state the rod and the barn move together, and the edges of the rod stick out of the barn symmetrically on both sides.

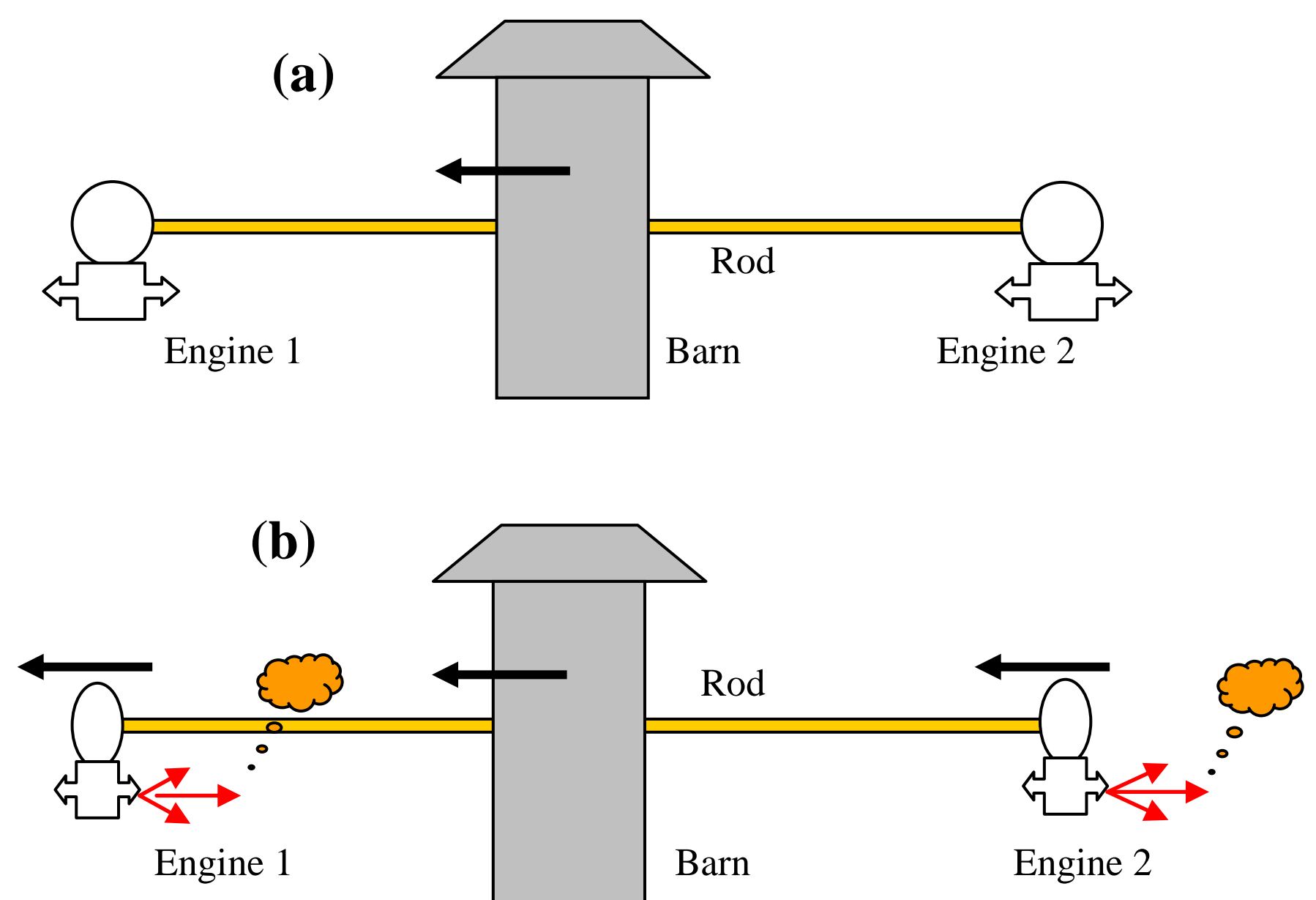

**Fig. 6**. Boosting all parts of the rod simultaneously in B. Assuming rod deformable, we represent it by two end masses with the attached engines.
(a) Right before the boost, the rod is stationary in B; (b) Right after boost, the rod is instantly accelerated to the left without changing its length.

In this situation, Bob and Alice exchange their roles. In case (I), Alice had observed the moving rod *stopped* with its length preserved. In case (II), Bob *boosts* his rod from rest so as to preserve its length. However, as we had already suspected, this does not conserve the *proper* length either. The same argument as (I) shows that the rod gets again *physically deformed* – but now its *proper* length *increases* by a factor of $\gamma(V)$:

$$L_0 \to \tilde{L}_0 = \gamma(V) L_0 \qquad (7)$$

This result follows directly from conservation of rod's length in B and Lorentz-contraction rule: since the rod is now *moving*, the rule implies (7). This is also consistent with Alice's observations, although her records are different from the picture observed in B. The boosting forces are *not simultaneous* in A (Fig. 7). The left engine starts first and stops the mass. The right-end mass keeps on moving and thus *extends* the rod. The engine at the right fires *after* passing by Alice. The moment it stops, both masses are at rest positioned symmetrically relative to the barn and are distance $\tilde{L}_0$ apart.

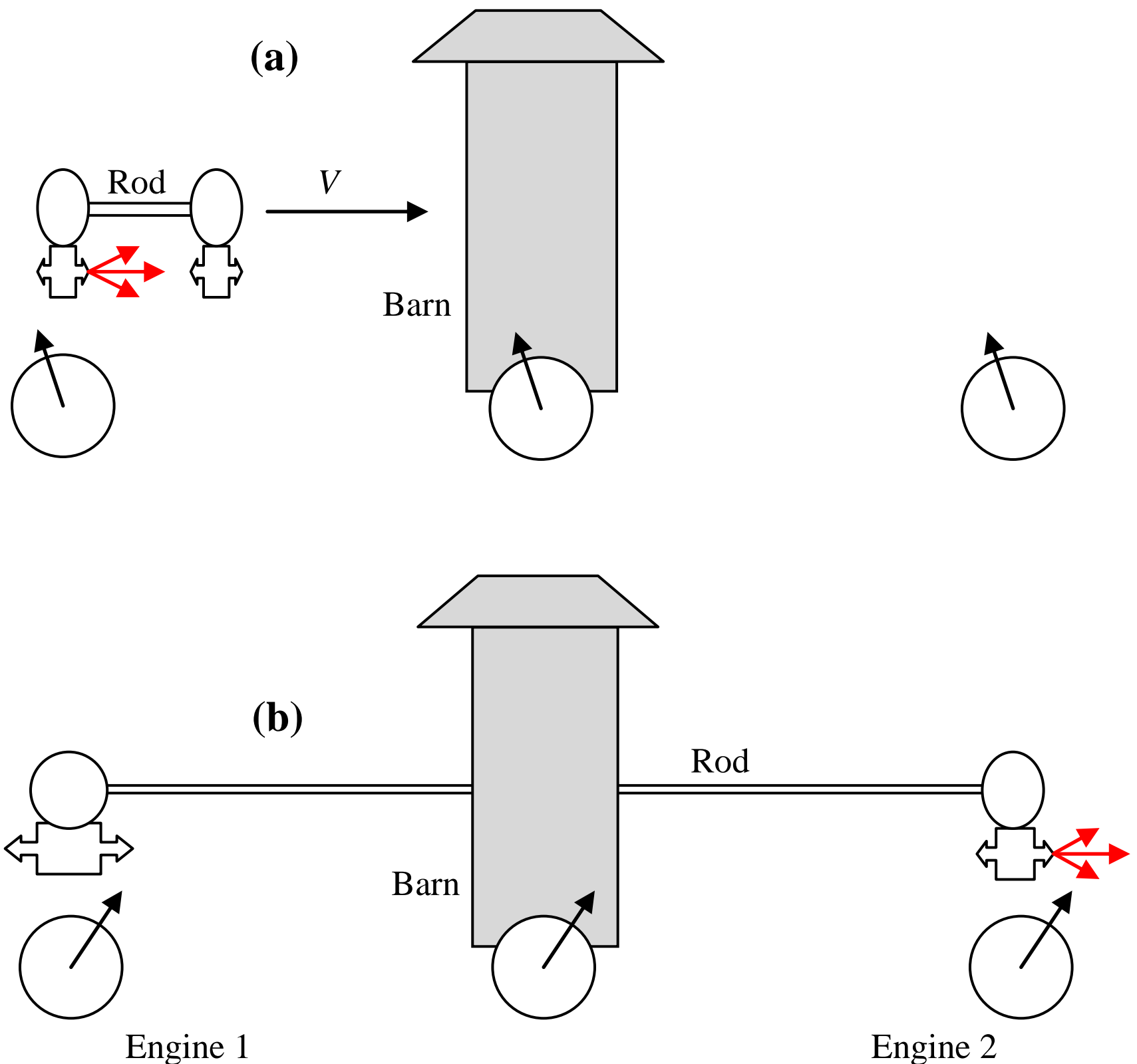

**Fig. 7**. The same process as in Fig. 6, but observed in A (not to scale).

(a) The initial moment of stopping the rod in Alice's RF. The left engine stops the left edge of the rod.

(b) The final moment: the right engine stops the right end.

Between the moments (a) and (b), Alice observes the rod being stretched from $L_0 / \gamma(v)$ to $\gamma(v) L_0$.

The quantitative description of the process in Alice's frame [18] also shows that if the rod is represented by a row of equidistant point masses, each with its own engine, and all the engines fire simultaneously to boost the rod without changing its length in B, then Alice observes a succession of consecutive flashes of the engines, each stopping its respective mass, and the pulse of flashes will run from the rear to the front of the rod with the same speed as in Eq. (4). Again, the superluminal pulse does not violate any laws, because it is not a motion of a *single* object.

Thus, simultaneous application of forces in B for preserving the length fails to preserve the *proper* length. But, contrary to the previous case, the rod now undergoes stretch.

How will this result change if there are inner forces? Let the end masses be connected by a relaxed spring. Right after the engines stop, the spring (which is now stretched!) will start contracting. The system begins to vibrate around its equilibrium size corresponding to the relaxed spring. This size is the rod's *initial* proper length. The vibrations will not generally be symmetric with respect to the relaxed configuration because the stretch may exceed the proper length (assuming unrestricted elasticity). If we still try to approximate deformation by Hooke's law, we will now need two effective spring constants, $k_c$ for compression and $k_s$ for stretch, with $k_c \to k_s \to k$ for small deformations, that is, for $\beta \equiv (V/c) \ll 1$. The potential energy gained by the rod in its transition from A to B can now be estimated as

$$U \cong \frac{1}{2} k_s L_0^2 \left(\gamma(V) - 1\right)^2 \tag{8}$$

Alternatively, we can from the very beginning introduce deformational potential energy $U(x)$ as the primary characteristic of the rod's material, with $x > 0$ for stretch and $x < 0$ for compression. Then (8) can be written as

$$U = U\left(x(V)\right), \quad x(V) = \left(\gamma(V) - 1\right) L_0 \tag{8a}$$

It follows that for an extended rod accelerated from rest under the above conditions, not only its proper length, but also its rest mass increases as it does in case (6), by the amount $\Delta M_0 = U\left(x(V)\right)/c^2$. Accordingly, more energy will be required to boost an object following this accelerating program. On the atomic level, the program II as seen from B and A, is shown in Fig. 8.

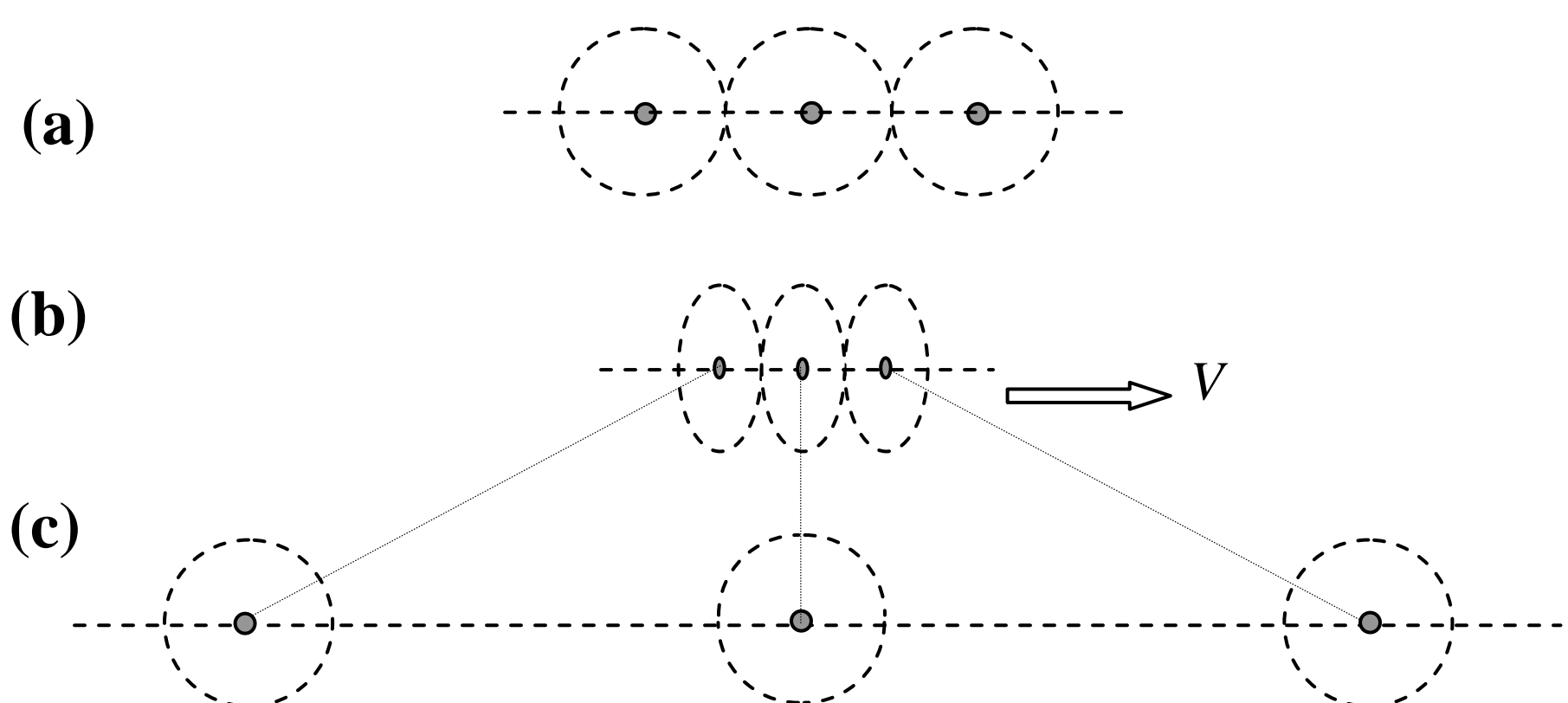

**Fig. 8**. The same as in Fig. 5, but now for case II
(a) In frame B; (b) In frame A before the system is stopped
(c) In frame A right after the stop. All the atoms along the rod are now mutually attracted

If the vibrations ensuing after removal of external forces are damped, the system will ultimately relax to its natural size, thus restoring its initial proper length and rest mass (the latter will reduce to its initial value due to dissipation of energy (8a)). Here the inner structure is again, as in case (I), the keeper of the initial shape. If, however, there remain some external forces maintaining the original stretch, then the object will remain after the boost in a physically deformed state with increased rest mass and its *proper* length extended by a factor of $\gamma\,(V)$. As we will see later, such forces appear in rotational motion of a ring or disk of a fixed radius.

(III) *Non-simultaneous braking*

There is more than one way to preserve the proper length in accelerated motion of a body.

One of them (incremental accelerating procedure) is described in [19]. A more straightforward (and accordingly, more violent!) solution is to just stop the trailing edge of the rod at the moment $t_1$ when its instant position is $x_1 = -L_0/2$, and the leading edge at the moment $t_2$ when its instant position is $x_2 = L_0/2$ [18]. This procedure insures conservation of proper length. In the process, the rod extends from $l_0$ to $L_0$ in A and contracts from $L_0$ to $l_0$ in B. If, instead of only end points, we represent the rod as a row of equidistant identical masses, stopped each by its individual engine one after another in equal short time intervals, starting at $t_1$ and ending at $t_2$, then Alice would observe the pulse of engine flashes rushing from rear to front of the stopping rod. The pulse separates two parts of the rod – one stopped and another still moving. The speed of the separation point is

$$u = \frac{x_2 - x_1}{t_2 - t_1} = \frac{L_0}{t_2 - t_1} = \frac{V}{1-\gamma^{-1}} = \frac{c^2}{V}\left(1+\frac{1}{\gamma}\right). \qquad (9)$$

Again, the separation point moves faster than light. In the end Alice sees the rod stopped in the position shown in Fig. 1. As mentioned, in the process she observed the rod being stretched from its Lorentz-contracted length $L_0/\gamma(V)$ to its proper length $L_0$. But she could only achieve this by stopping different parts of the rod *at different moments of her time*.

Bob observes the same process in reversed order: the moment $t_1'$ is *later* than $t_2'$. This does not contradict causality since the interval between the corresponding events is space-like and accordingly they are not in "cause and effect" relation. The pulse observed in B has the same speed (9) but in the opposite direction. It is an expansion pulse moving to the right in A, and a compression pulse moving to the left in B.

According to Bob, the rod shrinks from its proper length $L_0$ down to the Lorentz-contracted length $L_0/\gamma(V)$. But since the rod is now moving with speed $V$ relative to Bob, he concludes that its proper length measured by Alice is the same as the one originally measured by him – it is conserved. In either frame, the rod changes in length to preserve its *proper* length! It seems crazy but it is not. It would be self-contradiction for a uniformly moving or stationary rod in one inertial RF but has no conflict in case of its *acceleration.* The proper length here remains the same because the considered process of length change is relative. What is observed as compression by Bob is observed as stretching by Alice. One and the same system appears here to evolve in the opposite directions when viewed from two different RF. The *direction of evolution* of an accelerated object can be a relative property when the beginning and end of spatially evolving process in it have opposite ordering in these frames. Again, this does not contradict causality since the corresponding events in different parts of an object form the space-like intervals. The inner evolution of each single particle along its world line has the same direction for all observers.

(IV) *Generalization to arbitrary forces*

The situations considered above are thought experiments used to simplify the discussion. All local boosts were instantaneous due to infinite forces acting infinitesimally short time. But all outcomes actually depend on *impulse of force* $\Delta p = f\Delta t$ rather than on force itself. The definition of impulse holds for $f \to \infty$ and $\Delta t \to 0$ as long as their product remains finite. This justifies our model with infinite and instantaneous forces. Since the length $L_0 / \gamma(V)$ of moving rod depends on its velocity and thereby its final momentum, the result is the same as for finite forces lasting appropriate finite time to impart momentum $\Delta p$ .

We can restate it in the language of Lorentz transformations. Represent a rod by two equal end masses and accelerate it by imparting both masses with equal amount of momentum.

Consider two different acceleration procedures. In the first one both masses are hit simultaneously in their initial rest frame B by infinite instant forces parallel to the connecting spring. As was shown above, in the final state both masses (and corresponding rod) will move relative to B with a speed *V*, and the separation between them will remain the same as before. The length of the rod as measured in B will not change. But since the rod is now *moving* relative to B, it is Lorentz- contracted, which means that its *proper* length measured in its *new* rest frame A is now greater than it was before the boost by a factor of $\gamma(V)$ . The reason for change in proper length (the actual physical deformation) is relativity of simultaneity. Since the rod was initially moving in A, the applied forces were not simultaneous there: the trailing mass was stopped earlier than the leading mass, which resulted in extension of the rod.

In the second procedure both masses are subject to equal *finite* forces $f_1 = f_2$ , which start simultaneously in B and act during equal time intervals $\Delta t_1 = \Delta t_2 = \Delta t$ . Both masses travel equal distances $\Delta l_1 = \Delta l_2 = \Delta l$ during this time interval, so after procedure the rod flies through B at speed *V* with its length unchanged. But since it is now *moving*, it is Lorentz-contracted, which implies that its *proper* length (measured in its new rest frame A) has increased by the same factor $\gamma(V)$. The physical reason for this is the same as mentioned above: the masses are initially moving in A, so the two forces acting synchronously in B are not synchronous in A. Their Lorentz-transformed values $f_1'$ and $f_2'$ as well as the respective time intervals $\Delta t_1'$ and $\Delta t_2'$ of their action remain equal in A due to linearity of the Lorentz transformations. But due to relativity of simultaneity, they are *not* entirely overlapped in time as are their counterparts in B.

The force on the trailing mass starts earlier than does the force on the leading mass; and it ends earlier than does the force on the leading mass. Therefore, even though both masses travel the same distance $\Delta l_1' = \Delta l_2' = \Delta l'$ before their consecutive stops, the leading mass is stopped later than the trailing mass. This stretches the connecting spring, that is, extends the rod.

Thus, the second procedure produces *the same result* as the first one. The only difference is that after the second procedure the rod as a whole will be shifted by a distance $\Delta l$ in B or $\Delta l'$ in A as compared to its respective position in the first procedure. Since we are interested only in the *length* of the rod measured in two frames, rather than in its shift, this difference is immaterial, and one can use the assumption of the infinite instantaneous forces as was done in the previous sections. The second procedure is more realistic, but it tends to obscure the underlying physics behind additional mathematical details. And the underlying physics is that *an accelerated object does not generally conserve its proper length*. As shown above, the proper length conserves only under certain conditions; for instance, the external forces in case I must be slowly removed after completing the acceleration program. The case III shows that the procedure described in [19], while being the "least violent", is not the only one possible.

In the next section we use our approach for discussing rotational motions and the related Ehrenfest paradox.

## 2. The Ehrenfest paradox

Here we analyze the Lorentz contraction in rotational motions. The analysis will demonstrate the *physical mechanism* of deformation of a spinning disk and, accordingly, the *dynamical origin* of its specific geometry. This will explain the Ehrenfest paradox [29-33].

In contrast with situation in Sec. 1, now Alice's and Bob's frames are not equivalent within SR. Alice's frame is inertial while Bob's one is not. We denote them as $S$ and $S'$, respectively. Both observers now use cylindrical coordinates in the plane of the disk.

The "paradox" appears when we consider the rim of a spinning disk and try to find its proper length by the rules of SR. We assume that the rule (1) is applicable to the length of a circle uniformly rotating in its plane. This assumption can be backed by a simple argument. Even though the spinning disk does not form an inertial RF, each of its sufficiently small area elements can be considered as a part of co-moving inertial RF in which this specific element is currently at rest. Then this element as observed in the Lab must be Lorentz-contracted along the direction of its motion. And since this holds for each element, we expect the whole rim to be Lorentz-contracted. But we seem to run into trouble when describing this quantitatively.

Consider a disk spinning in its plane with angular velocity $\Omega$. All length elements $dr$ of its radius $R$ are perpendicular to the instant direction of their motion. Therefore the value of $R$ is the same for both observers.

But it is not that simple for the rim. Each element of the rim instantly passing by Alice has a speed $V = \Omega R$ and is Lorentz-contracted. Accordingly, the whole circumference $L$ measured as the sum of its increments is less than its proper length $\Lambda$ by the corresponding Lorentz factor:

$$L = \frac{\Lambda}{\gamma(V)}, \qquad \text{or} \quad \Lambda = L\gamma(V) = 2\pi R\, \gamma(V) \tag{10}$$

But this seems impossible since the spinning circumference remains congruent with its stationary counterpart. Such congruence (coincidence) defines geometric equality. Therefore there must be just

$$\Lambda = L = 2\pi R \tag{11}$$

These two conflicting statements make mathematical formulation of the Ehrenfest paradox. The paradox originates from the assumption of Euclidean geometry on rotating disk. But this assumption is wrong, and the whole space-time, while remaining Lorentzian everywhere on the disk, has a peculiar topology not admitting single global time [17, 32-34]. The circumference of rotating disk measured by Alice *is* shorter than its *proper* circumference measured by Bob. It is equal to $2\pi R$ for Alice, thus satisfying the congruence, but is greater than $2\pi R$ for Bob as in Eq. (10)[1]. This means that a spinning disk is in a state of a complex deformation which renders its plane non-Euclidean. It is described by Lobachevsky's geometry with negative curvature; sometimes it is referred to as a hyperbolic geometry – by the name of a surface (one-folded hyperboloid of revolution) on which it is realized[1]. Its most essential features are that the sum of the angles of a triangle is less than $2\pi$, and the ratio of the length of a circle to its radius is greater than $2\pi$. A more detailed description can be found, e.g., in [17].

[1] As we will see later, the congruence *in the co-rotating frame* is satisfied in a more subtle way taking account of the fact that time in rotating systems is not single-valued.

Now we can discuss the connection between the dynamics of spinning disk and its geometry. To simplify, let us focus on the rim of the disk and call it the ring. Being a part of the disk, it is rotating with angular velocity $\Omega$ about its symmetry axis perpendicular to its plane.

As in Sec. 1, we will consider the *transition* from the stationary to a spinning state. We start from a stationary ring, bring it to rotation and watch how this transition affects the geometry.

Start with two congruent rings both stationary in *S*. One of them is scheduled for rotational boost and has each of its length elements $dl = R\,d\varphi$ connected with the ring's center by a rigid spoke to preserve its radius *R*, so that it looks like a wheel of a bicycle. The scheduled operations include:

(1) The rotational boost of the wheel from *S* to $S'$;

(2) Length measurement of both rings in *S* and $S'$.

It is natural to mark the stationary ring by the same notation *S* as Alice's Lab, while the *boosted* wheel by notation $S'$ of Bob's frame. In order to maintain congruence between *S* and $S'$ we boost the wheel by applying equal tangential forces to its equal parts *dl simultaneously in S*. This, as we know, preserves the size of each element *dl* and thus the whole length of boosted wheel *in S*, even though the wheel is now rotating with linear velocity $v_R = \Omega R$. But, while remaining congruent with *S*, ring $S'$ is now Lorentz-contracted, which means that its *proper* length must increase due to some stretching deformation. To trace out its origin, turn to Bob's records in $S'$. If $S'$ rotates counter-clockwise relative to *S*, the *wheal before the boost* is spinning clockwise with respect to $S'$ (Fig. 9). Each of its increments is initially Lorentz-contracted as

$$dl'_{In} = \frac{dl}{\gamma(V)} \qquad (12)$$

While all increments are boosted by Alice simultaneously, the corresponding local boosts (which are stops in $S'$) are *not* simultaneous there. According to the Lorentz transformation with $dt = 0$, the moments of stops of $dl$'s edges are separated by the time interval

$$dt' = -\gamma(V)\frac{V}{c^2}dl \qquad (13)$$

The minus sign here indicates that at one moment of Alice's time, Bob's clock instantly coincident with trailing edge *N* of the boosted segment $MN = dl$ reads an earlier time than his clock instantly coincident with the leading edge *M*. Therefore when Bob sees edge *N* stopped relative to him, edge *M* keeps on moving with speed *V* to be stopped later in accordance with (13). This stretches $dl$ – without changing the angle $d\varphi$ it subtends! The amount of stretch is

[1] A non-Eucledian geometry on a curved 2-dimensional surface can be realized in Euclidean 3-dimensional space. Similarly, the non-Euclidean spatial geometry on rotating disk does not affect geometry of the embracing space-time which remains Lorentzian.

$$\delta' = -V\,dt' = \gamma(V)\frac{V^2}{c^2}\,dl \qquad (14)$$

As in Sec. 1, we can represent the element by two equal masses at its two edges connected by a spring. The spring will keep on stretching until the opposite mass will be instantly stopped, thus transferring the entire element to $S'$. Its resulting new length in $S'$ is the sum of (12) and the amount of stretch:

$$dl'_{Fin} = dl'_{In} + \delta' = \gamma(V)\,dl \qquad (15)$$

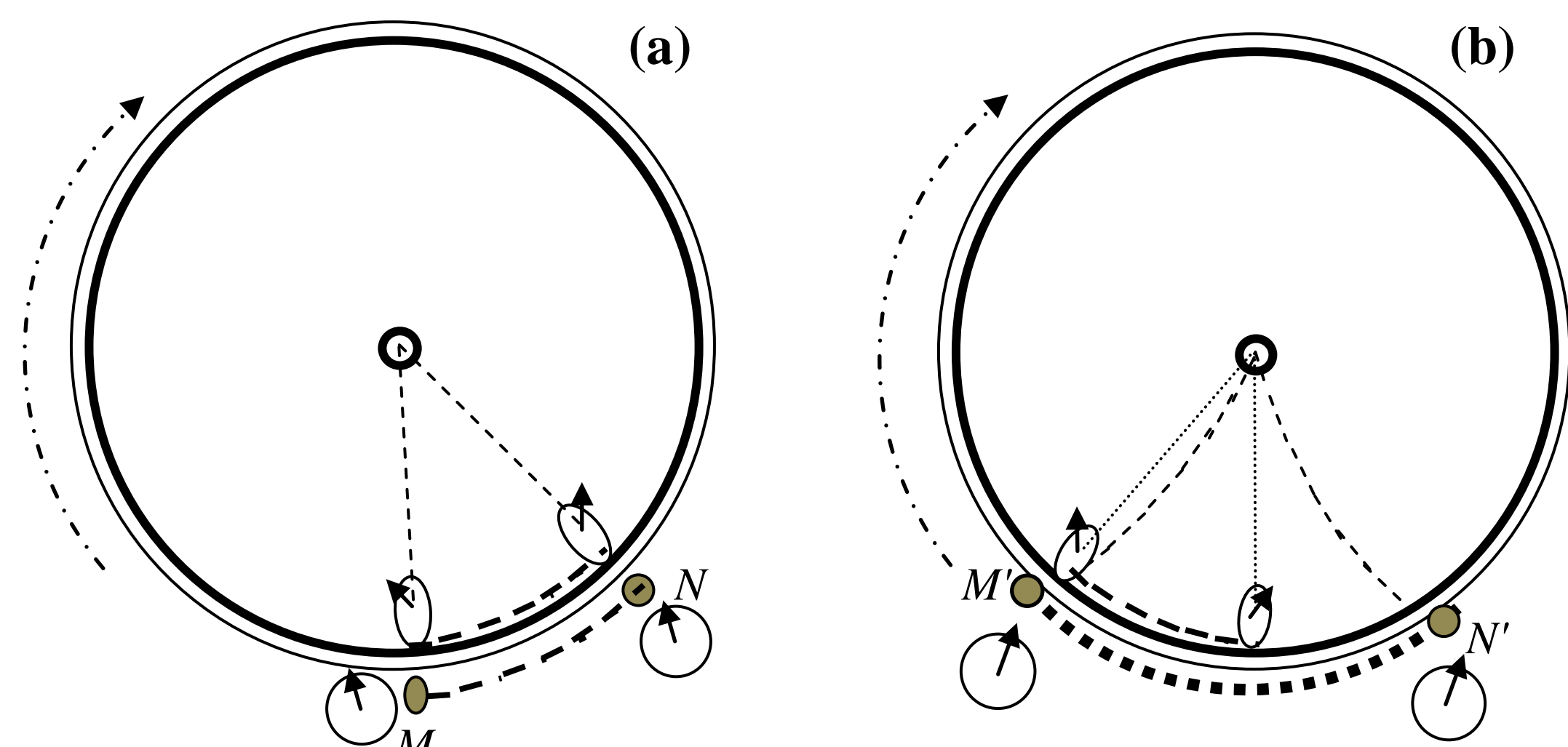

**Fig. 9** The rotational boost of ring's segment *MN* from *S* to as seen by Bob. It is shown as the external ark subtending central angle $\varphi$. Initially it is rotating clockwise in $S'$.
(a) Initial moment ($t' = -1$) of Bob's time. *MN* is instantly coincident with element $dl$ (both shown as dashed segments) but edge *N* is instantly stopped while *M* keeps orbiting clockwise.
(b) Final moment ($t' = +1$). Edge *M* and thereby the whole segment stopped in $S'$.The resulting segment $M'N'$ (now shown square-dotted) is accordingly extended (without increasing the subtended angle $d\varphi$!), and the same holds for the whole ring after its transfer from *S* to $S'$. But these features cannot be shown adequately on the Euclidean plane of the figure

Since the element *dl* was chosen arbitrarily, this result holds for all other elements, and thereby for the whole ring. This proves result (10). Formally we can obtain (10) by integrating (15):

$$\Lambda(R) = \oint dl'_{Fin} = \gamma(V)\oint dl = \gamma(V)\,L = 2\pi R\,\gamma(V) \qquad (16)$$

This result shows again that the so-called kinematic (or, as many say, geometric) effect has a dynamic origin – a non-simultaneous action of boosting forces in $S'$.

Both – Alice and Bob – come to the same conclusion that the rings $S$ and $S'$, while having the same radius $R$ and being congruent, have different proper lengths. The ring $S$ is in the relaxed state and has the "normal" length $L = 2\pi R$; the $S'$-ring is in a deformed state: when measured in $S'$, it is longer than $L$ as described by Eq. (16). Both observers agree on the results, but give different explanations for it. For Alice, since $S'$ is congruent with $S$ despite the Lorentz-contraction, its *proper* length must exceed $L$ by the Lorentz-factor. For Bob, the ring $S'$ is longer than $S$ ($\Lambda > L$) by the same factor, because it had been stretched due to non-simultaneous application of boosting (that is, stopping!) forces. The fact that $\Lambda$ is greater than $2\pi R$ is explained by Bob as a manifestation of the non-Euclidian (Lobachevsky's) geometry in a rotating system. In contrast with case in Sec. 1 when *both* frames A and B are inertial and the deformed system requires external forces to maintain the post-transitional proper length, here the proper length $\Lambda$ of Bob's ring remains *self-sustained* since the ring is closed loop.

Some treatments [35-37] explain the discussed "paradox" by stating that SR is not applicable in $S'$. In reality, SR can be consistently used in non-inertial RF-s if we do it locally [7, 18]. All obtained results are in agreement with observations and, as shown above, have simple physical explanation within its framework.

But this becomes more complicated when we consider the whole problem the other way around. Namely, assuming that Bob can locally apply the rules of SR to $S'$, can he claim that the *proper* length $L$ of Alice's ring must be greater than $\Lambda$ in order to insure congruence with his ring?

The answer to this question may seem "Yes". Directly measured length of $S$ should equal $\Lambda$ due to permanent congruency of both rings. But if SR works in $S'$ and each element of $S$ is Lorentz-contracted it must be shorter than element's *proper* length $dl$ measured by Alice. This holds for all elements and thereby for the whole ring. Bob thus might conclude that Alice's ring's *proper* length $L$ is greater than $\Lambda$:

$$L = \gamma(V)\Lambda > \Lambda , \qquad (17)$$

in flat contradiction with (16). Thus, there are two different lines of reasoning leading to two conflicting conclusions.

This paradox can still be resolved within the framework of SR.

Start with the procedure of the length measurement of $S$ by Bob. The crucial requirement is that leading and trailing edges of a *moving* object *must be marked simultaneously* in RF where we measure its length. Since the object (ring) is a closed loop, its leading and trailing edges coincide. Therefore the requirement that their positions be marked at the same moment of $S'$-time seems to be satisfied automatically.

That would be true in *inertial* RF. But in $S'$, the situation is more subtle.

At one moment in $S$, a pair of *synchronized* clocks on ring $S'$ reads different times. Following succession of the ring's clocks until we return to the original one, we realize that it should read two different times at once [18, 33]. Integrating (13) along the whole ring gives the corresponding time lag

$$\Delta t' = -\gamma(V)\frac{V}{c^2}\oint d\,l = -\gamma(V)\frac{V}{c^2}L \qquad (18)$$

or, since $V = \Omega R$ and $L = 2\pi R$:

$$\Delta t' = -2\pi\gamma(V)\frac{\Omega}{c^2}R^2 = -2\frac{\gamma(V)}{c^2}\Omega A \qquad (19)$$

The time lag is proportional to Ω and to the area $A = \pi R^2$ enclosed by the loop. This can be generalized to the loops of an arbitrary shape [7, 34].
Reiterating will attribute a set of different equidistant moments to the same event:

$$t' \to t'_m \equiv t' + m\,\Delta t' \;, \quad |m| = 0, 1, 2, \ldots \qquad (20)$$

The origin of time lag can be visualized if we apply the rules of Minkowski geometry to rotating frames. Temporal and spatial axes $x'$, $ct'$ of *inertial* frame B moving at a speed *V* in the *x*-direction of A are tilted towards each other by the angle $\theta = Arc\tan\left(V / c\right)$ when represented in A (Fig. 10a). But for *rotational* motion, the *x*-axis converts into a circle of chosen radius *R*. Adding temporal dimension leads to the corresponding cylindrical surface (Fig. 10b). The result can be envisioned in frame *S*. Since $S'$ is rotating, the lines $x'$ and $ct'$ form the respective helixes with different tilts. The helix $x'$ is the line of simultaneity in S′, or, equivalently, the world-line of a fictitious particle moving infinitely fast along the ring in $S'$; and the helix OO′ (with only the front part shown) is the world-line of a photon moving in a circle along the ring. As in Fig.10a, the angles made by $ct'$ and $x'$ with axis $ct$ are $\theta$ and $\left(\pi / 2\right) - \theta$, respectively. They have different pitches and therefore will have multiple intersections. This means that the same single event is associated with a set of different moments $t'_m$ of time in rotating frame. The interval of proper time $\Delta t'$ between two successive intersections is the time lag given by Eq-s (18, 19). The space and time in a rotating frame are literally intertwined as are the two helixes representing them in Fig. 10b.

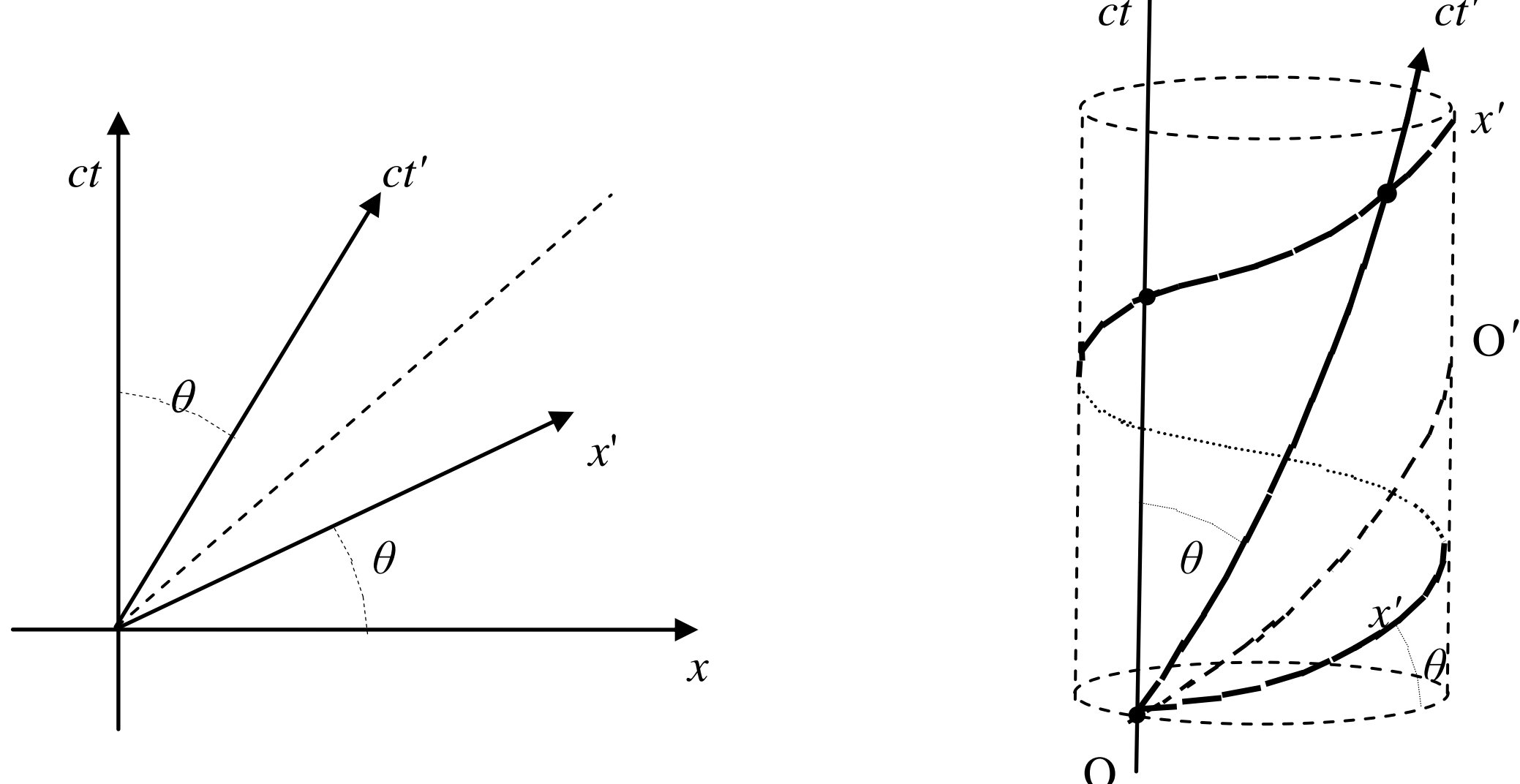

**Fig. 10** Space-time diagrams for translational and rotational motion
(a) Both – A and B – are inertial (b) $S$ is inertial, but $S'$ is rotating.

The time lag in a rotating system is manifest in an observable effect [17, 18, 36, 37]: a reference clock $C_R$ near Bob on the rim of disk $S'$ will record different circumnavigation times for two objects boosted from $C_R$ along the rim at the same local speed *v* but in the opposite senses. Call the object boosted in direction of rotation the "E-object", and the one boosted simultaneously in the opposite direction – the "W-object". After one circumnavigation, the W-object will return to $C_R$ earlier than the E-object. The corresponding proper times $\tau_W$ and $\tau_E$ read by $C_R$ at the moments of return are [17, 18, 38, 39]

$$\tau_E = t_E / \gamma(V) = \tau_v \left(1 + \frac{V v}{c^2}\right); \qquad (21a)$$

$$\tau_W = t_W / \gamma(V) = \tau_v \left(1 - \frac{V v}{c^2}\right) \qquad (21b)$$

Here $t_E$ and $t_W$ are the respective *coordinate times* read by the *S*-clocks, and

$$\tau_v \equiv \frac{\Lambda}{v} = \frac{2\pi R \gamma(V)}{v} \qquad (22)$$

is the characteristic time for traveling distance $\Lambda$ with speed *v* along a *non-closed* path.

Equations (21) were initially obtained for a special case of the Earth-circumnavigation and experimentally confirmed. But they are quite general and may even hold for superluminal *v* (such velocities, as mentioned before, do not by themselves contradict anything). Their quantitative description is consistent with results (18-20) and gives another view of the phenomenon.

For our purpose we consider two fictitious particles (tachyons [17, 18, 40-43]) launched simultaneously in the E- and W directions along the circumference with infinite speed *relative to* $C_R$. Accordingly, let $v \to \infty$ in Eq-s (21, 22). As mentioned above, the world-line of a particle with infinite velocity along certain direction in a given RF determines the corresponding simultaneity line along this direction. For a spinning disk, this world-line is coincident, e.g., with its rim, but in *S*, it is the respective helix $x'$ shown in Fig. 10b. And again, we see that it has multiple intersections with $ct'$.

Eq-s (21) at $v \to \infty$ give two corresponding moments of proper time of $C_R$ at such intersections:

$$\begin{aligned} \tau_E(\infty) &= \lim_{v \to \infty} \frac{\Lambda(V)}{v}\left(1 + \frac{Vv}{c^2}\right) = \frac{V}{c^2}\Lambda(V) \\ \tau_W(\infty) &= \lim_{v \to \infty} \frac{\Lambda(V)}{v}\left(1 - \frac{V v}{c^2}\right) = -\frac{V}{c^2}\Lambda(V) \end{aligned} \qquad (23)$$

Since $\Lambda(V) = \gamma(V) L$ and $\tau$ is the corresponding local time coordinate $t'$ in the rotating system, this is identical to expressions (18, 20) with $|m| = 1$ (one circumnavigation) for the time lag for the same event (intersection between the axes!). Indeed, once the fictitious particle is moving infinitely fast, it is expected to return immediately. But Eq-s (23) assign finite waiting time for return: it occurs *after* the moment of departure for the E-particle, and, even more bizarre, *before* the moment of departure (negative sign of $\tau_W(\infty)$ !) for the W-particle. Both these times are as legitimate as the zero moment $\tau = 0$, and are accordingly represented together with this moment, on the same footing, as the temporal characteristics of one event. The difference between them is the time lag discussed above.

The diagram in Fig. 10 prompts another derivation of the time lag lying at the heart of the Ehrenfest's paradox. The $x'$-line, by its original definition, is the line $ct' = 0$. But here it intersects the $ct'$- line at $ct' \neq 0$ as well. The intersection point is at the point $x' = 0$ of Bob's location. So one event at definite location has at least two distinct times $t' = 0$ and $t' \neq 0$.

We now find this alternative time $t'$ in different way than before. We use the relation between time $t'$ in $S'$ and time $t$ in Alice's frame. Once we find the moment $ct$ of Alice's time corresponding to intersection, we can use the relation $t \leftrightarrow t'$ to find the corresponding $ct'$.

Since Bob is at $x' = 0$ of $S'$, the relation is

$$c\Delta t = \gamma(V)\, c\Delta t' \tag{24}$$

Bob's world-line is described by

$$ct = \tan\left(\frac{\pi}{2} - \theta\right) x = \frac{R\varphi}{\tan\theta} = \frac{R\varphi}{\beta}, \quad \text{where } \beta \equiv \frac{V}{c} \tag{25}$$

The equation for $S'$-s simultaneity line is

$$ct = (\tan\theta)\, x = \beta R \varphi \tag{26}$$

At the point $t_1 = \Delta t$ of their (first) intersection we have

$$\frac{R}{\beta}\varphi = R\beta(\varphi + 2\pi) \tag{27}$$

(Actually, we can, instead of $2\pi$, add $2\pi m$ with integer $m$, since the process reiterates). Solving for $\varphi$, we find that this happens at

$$\varphi = 2\pi\beta^2\gamma^2(V) \tag{28}$$

Putting this into (25) gives

$$ct_1 = 2\pi R\beta\gamma^2(V), \tag{29}$$

so that, in view of (25) we finally obtain

$$c\Delta t' = \frac{c\Delta t}{\gamma(V)} = 2\pi R\beta\gamma(V) \tag{30}$$

or, since $\beta = V/c = \Omega R/c$,

$$\Delta t' = 2\gamma(V)\frac{\Omega}{c^2}A, \qquad (31)$$

where $A = \pi R^2$ is the (Euclidian!) area of the disk. This recovers Eq-s (18-20) at $m = -1$.

If $ct'$ is Bob's world-line, the intersection O′ of this line with $x'$ is an event of Bob's biography different from O, and it is accordingly assigned a later moment given by (31). But since it is connected with O by the world-line of a particle with $v = \infty$ in $S'$ (that is, by simultaneity line in this frame), it can be assigned the time $t' = 0$ as well. Both moments are equally legal. By reciprocity, the event O can also be assigned the earlier time equal to negative of (31). And by iteration, all multiple integers of (31) would be also legal. And since the choice of a reference event is arbitrary, the obtained conclusion holds for any event on the spinning disk (except for those at its center). Finally, since $c\,\Delta t'$ is the time elapsed after the zero moment from which we count time in this example, the obtained equation is just a special case (at $m = -1$) of the general formulas (18-20) for the time lag $\Delta t'$.

The absence of single global time in rotating frames is also evident in another consequence of Fig. 10b and general relations (21). It is in geometry of photon world lines. The world lines of the E- and W-photons, while being symmetric relative to $ct$, have different intersection points with $ct'$. Physically this means that if Bob uses a source emitting simultaneously two oppositely-directed photons along the rim of his disk, the photons return at different moments of his time. Setting in (21) $v = c$ gives

$$\tau_E = \frac{\Lambda}{c}\left(1 + \frac{V}{c}\right), \quad \tau_W = \frac{\Lambda}{c}\left(1 - \frac{V}{c}\right) \qquad (32)$$

so that

$$\Delta\tau \equiv \tau_E - \tau_W = 4\gamma(V)\frac{A\Omega}{c^2} \qquad (33)$$

If Bob decides to use such an experiment for measuring speed of light as the ratio of $\Lambda/\tau$, he will get different speeds for the E- and W- photons:

$$c_E = \frac{\Lambda}{\tau_E} = \frac{c}{1 + \frac{V}{c}} \quad \text{and} \quad c_W = \frac{\Lambda}{\tau_W} = \frac{c}{1 - \frac{V}{c}} \qquad (34)$$

where $c$ is the local speed of light. If Michelson would have performed such experiment on a rapidly rotating platform, he would "prove" the existence of ether with the "ether wind" swirling around the platform's center.

In reality, it only shows the peculiar topology of space-time in rotating frames, where the speed of light along a closed path is a totally different characteristic than its single-valued local speed.

Now we are in a position to return to the Ehrenfest paradox and complete its discussion. In order for the length measurement of ring $S$ to be instantaneous *in* $S'$, we need to account for the time lag (31). Thus, the total length of ring $S$ in $S'$ will consist of two contributions: its $A$-congruent length *and* the additional distance $V\,\Delta t'$ traveled by a point on the ring during the time lag $\Delta t'$:

$$L' = \Lambda + V\Delta t' = \Lambda - \frac{V^2}{c^2}\Lambda = \frac{\Lambda}{\gamma^2(V)} = \frac{L}{\gamma(V)} \quad (\neq \Lambda!) \qquad (35)$$

This is precisely the Lorentz-contracted length of *S* as measured in $S'$. This result is obtained only if: 1) Bob marks the position of a point on ring S at two different moments of his time separated by the time lag $\Delta t'$ and then 2) subtracts the distance between the two consecutive markings from Λ. This subtraction takes account of the fact that both moments of marking are equally legitimate temporal labels of one event in Bob's frame. Crazy as it may seem, such procedure constitutes a length measurement of a ring S in system $S'$ and gives an operational definition of measurement result. The procedure consistently applies the rules of SR and the obtained result (35) is also consistent with these rules. From this result Bob concludes that the proper length of ring *S* is $\gamma(V)L' = L$, in total agreement with Alice's record.

Thus, the $S'$-ring is Lorentz-contracted in *S* and has its proper length Λ greater than *L*; this result is in agreement with [32]. At the same time, the question asked by Bob whether the ring *S* is Lorentz-contracted in his frame $S'$, also has a positive answer. The ring *S*, while having its proper length *L* less than Λ, is Lorentz-contracted in $S'$. The two apparently contradicting statements are reconciled by taking into account the time lag.

The whole argument may seem somewhat superficial, but it is logically impeccable and just reflects totally unusual character of the described phenomenon. Nothing seems to be more remote from the domain of our intuition than the notion of the two distinct but equally legitimate moments of proper time characterizing a given event. There is no such thing as one single moment of time for all space in a rotating system. And the considered effect illustrates how the formalism of SR works in accelerated RF and gives a coherent description of the whole process consistent for all observers.

Thus, the geometry in a rotating system is affected on both – local and global scales. The former was shown to be the result of dynamics of rotational boost. The latter is manifest in experiments involving closed loops.

In a disk whose particles are bound together by internal forces this result – purely circumferential stretch – causes, among other things, the appearance of additional potential energy $U(R)$ associated with circumferential elastic deformation.

For a quantitative description consider, by analogy with Eq-s (8) for translational motion, the potential energy of elastic deformation of the whole disk. Substitute the circumference of radius *R* in Eq-s (16, 17) with an arbitrary annulus of a radius *r*, $0 \le r \le R$. Introduce the linear energy density along a radial direction:

$$\chi(r) \equiv \frac{d\left(U(r)\right)}{dr} \equiv \chi\left(\Delta L(v_r)\right), \qquad \Delta L(v_r) \equiv \Lambda(v_r) - L(r) = 2\pi r\left(\gamma(v_r) - 1\right) \quad (36)$$

Then

$$U(R) = \int_0^R \chi(r)\,dr, \quad \text{and} \quad \Delta M_0' = U(R)/c^2 \qquad (37)$$

Here $\Delta M_0'$ is the corresponding increase of the rest mass of the spinning disk in the co-rotating frame S′. In contrast with the translational motion, in which the boosted elastic object eventually restores both – its *initial* proper length and the rest mass after gradual removal of the external forces, a spinning object *retains* its *new* proper length and the rest mass after a *rotational* boost. If we have a system of identical equidistant masses arranged in a circle, and each mass is connected to the center with a sufficiently rigid spoke, then the circumferential springs connecting the masses, even though stretched during the transition $S \to S'$, form a system that remains self-sustained after the removal of the external forces.

The obtained result gives us a criterion for a possible experimental observation of the described effect. Consider first a separate annulus of radius *r* with the linear speed $v_r = \Omega r$. If one of its springs is cut, the corresponding adjacent masses will start moving apart along the circle until all the masses eventually cluster over the ark of length $\tilde{L}(r) = 2\pi r / \gamma(v_r)$ in the Lab frame S, with the center of the ark opposite to the cut spring.

Alternatively, one of the springs must be the first to spontaneously snap at sufficiently rapid rotation, with the same result. This is purely relativistic effect, which can in principle be observed under the corresponding conditions. Another manifestation of this effect is the increase of the amount of energy necessary for the boost as compared with that necessary for the similar boost of the system of non-interacting masses of the same shape. Using the definition (36), this additional energy required in S for the boost can be evaluated as[1]

$$\Delta \mathcal{E} = \int_0^R \gamma(v_r)\, \chi(r)\, dr \qquad (38)$$

The specific form of function $\chi(r)$ cannot be derived from the basic principles only, and depends on physical structure of the disk.

## Conclusion

We can now summarize the results of all considered thought experiments.

1. The relativistic kinematics of accelerated objects cannot be separated from the dynamics. A change in motion of particles constituting an object, changes the structure of their fields and thereby the shape of the object after acceleration [18, 26-28]. The outcome of an acceleration procedure depends on both – details of the process and *physical structure* of the object.
2. The size of an accelerated object cannot be uniquely determined by (1), because the object generally cannot even be assigned a *constant* proper length.
3. The concept of deformation in relativistic mechanics is more subtle than in classical physics. Its two intimately linked characteristics – geometric shape *and* physical structure – are not rigidly correlated. The former is a relative attribute and therefore may not manifest itself in certain RF in a deformed state, and vice versa, be manifest in some RF in a deformation-free state. A uniformly moving rod is technically deformed (length contraction), but physically it is deformation-free, which becomes evident in its rest frame. A uniformly rotating ring, while *retaining its circumference length* $L = 2\pi R$, is physically deformed (circumferentially stretched at fixed *R*), which becomes evident in the co-rotating RF.

---

[1] We do not consider here the additional potential energy associated with radial deformation of the spokes. Even though this deformation is assumed to be negligibly small here, the corresponding energy may be comparable to or even exceed that associated with the circumferential deformation. We do not include it merely because it does not present interest here, since it is not associated with any change of geometry.

4. Simultaneous application of equal braking forces to equal parts of a moving rod tends to compress the rod in its original rest frame, thus decreasing its proper length.
5. Simultaneous application of equal accelerating forces to equal parts of a stationary rod tends to stretch the rod in its final rest frame, thus increasing its proper length.
6. If one stops a rod by applying non-simultaneous local forces timed specially to preserve its proper length, it tends to stretch the rod in the RF where it had originally moved, and compress it in RF where it had originally rested.
7. If an object is boosted from one inertial RF to another, the binding *internal fields and interactions* within the object tend to ultimately restore its proper length perturbed by the boost.
8. In contrast, in a rotational boost, an object undergoes physical deformation lasting permanently after the boost and becoming one of the characteristics of its spinning state.
9. Study of *transitions* between different states of motion provides a deeper insight into the nature of the so-called kinematic effects. In particular, the dynamical aspect of the Lobachevsky's geometry in a rotating system is manifest in the increase of the system's rest mass. This is associated with an additional energy input necessary for boosting such a system, apart from the energy going into increase of relativistic mass of its constituting particles.

**Acknowledgements**
I want to thank David Green for turning my attention to the barn and rod paradox.